\documentclass{article}
\usepackage{spconf,amsmath,graphicx}

\usepackage{hyperref}
\usepackage[utf8]{inputenc} 
\usepackage[T1]{fontenc}    
\usepackage{url}            
\usepackage{booktabs}       
\usepackage{amsfonts}       
\usepackage{nicefrac}       
\usepackage{microtype}      
\usepackage{enumitem}
\usepackage{algorithm}
\usepackage{algpseudocode}
\usepackage{wrapfig}
\usepackage{subcaption}
\usepackage{multirow}
\usepackage{color}
\usepackage{amsmath,bm}

\title{AdaSpeech 2: Adaptive Text to Speech with Untranscribed Data}
\name{Yuzi Yan$^1$, Xu Tan$^2$, Bohan Li$^3$, Tao Qin$^2$, Sheng Zhao$^3$,  Yuan Shen$^1$, Tie-Yan Liu$^2$}
\address{$^1$Department of Electronic Engineering, Tsinghua University, Beijing, China\\ $^2$Microsoft Research Asia, $^3$Microsoft Azure Speech}
\begin{document}
%
\maketitle
\begin{abstract}
Text to speech (TTS) is widely used to synthesize personal voice for a target speaker, where a well-trained source TTS model is fine-tuned with few paired adaptation data (speech and its transcripts) on this target speaker. However, in many scenarios, only untranscribed speech data is available for adaptation, which brings challenges to the previous TTS adaptation pipelines (e.g., AdaSpeech). In this paper, we develop AdaSpeech 2, an adaptive TTS system that only leverages untranscribed speech data for adaptation. Specifically, we introduce a mel-spectrogram encoder to a well-trained TTS model to conduct speech reconstruction, and at the same time constrain the output sequence of the mel-spectrogram encoder to be close to that of the original phoneme encoder. In adaptation, we use untranscribed speech data for speech reconstruction and only fine-tune the TTS decoder. AdaSpeech 2 has two advantages: 1) Pluggable: our system can be easily applied to existing trained TTS models without re-training. 2) Effective: our system achieves on-par voice quality with the transcribed TTS adaptation (e.g., AdaSpeech) with the same amount of untranscribed data, and achieves better voice quality than previous untranscribed adaptation methods\footnote{Synthesized speech samples can be found at \url{https://speechresearch.github.io/adaspeech2/}}\footnote{This work was done while the first author was interning at Microsoft. Correspondence to: Xu Tan.}\footnote{Accepted by ICASSP 2021}.

\end{abstract}  
\section{Introduction}
\label{sec:intro}

Text to speech (TTS) adaptation has been widely adopted to synthesize personalized voice for a target speaker. The typical TTS adaptation pipeline~\cite{chen2018sample,moss2020boffin,arik2018neural,jia2018transfer,cooper2020zero,li2017deep,wan2018generalized,chen2021adaspeech} usually fine-tunes a well-trained multi-speaker TTS model~\cite{wang2017tacotron,ren2020fastspeech,chen2020multispeech} (source model) with few adaptation data, and achieves good similarity with ground-truth human voice and high voice quality.

Most previous works~\cite{chen2018sample,arik2018neural,jia2018transfer,chen2021adaspeech} on TTS adaptation require paired data (speech and its transcripts), such as AdaSpeech~\cite{chen2021adaspeech}. However, paired data is harder to obtain than untranscribed speech data, which can be broadly accessible through conversations, talks, public speech, etc. Therefore, leveraging untranscribed speech data for TTS adaptation can greatly extend its application scenarios. A basic method is to first use an ASR system to transcribe the speech into text and conduct adaptation as in previous works. However, this additional ASR system may not be available in some scenarios, and its recognition accuracy is not high enough which will generate incorrect transcripts and affect the adaptation. There indeed exist some works \cite{luong2020nautilus,taigman2017voiceloop,Luong2019A} using untranscribed data for TTS adaptation, which requires joint training of TTS pipeline and the modules used for adaptation. This joint training is not pluggable and restricts their method to be extended to other common TTS models.

In this paper, we propose AdaSpeech 2 that leverages untranscribed speech data for adaptation. We use the basic structure of AdaSpeech~\cite{chen2021adaspeech} as our source TTS model backbone, which contains a phoneme encoder and a mel-spectrogram decoder. To enable untranscribed speech adaptation, we additionally introduce a mel-spectrogram encoder into the well-trained source TTS model for speech reconstruction together with the mel decoder\footnote{In this paper, mel-spectrogram is sometimes abbreviated to mel for simplicity.}. At the same time, we constrain the output sequence of the mel encoder to be close to that of the phoneme encoder with an L2 loss. After training this additional mel encoder, we use untranscribed speech data for speech reconstruction and only fine-tune a part of the model parameters in the mel decoder following ~\cite{chen2021adaspeech}, while the mel encoder and the phoneme encoder remain unchanged. In inference, the fine-tuned mel decoder along with the unchanged phoneme encoder form a TTS pipeline that can synthesize custom voice for target speaker.

AdaSpeech 2 has two main advantages. 1) Pluggable. As the essence of our method is to add an additional encoder to a trained TTS pipeline, it can be easily applied to existing TTS models without re-training, improving reusability and extendibility. 2) Effective. It achieves on-par voice quality with the transcribed TTS adaptation, with the same amount of untranscribed data. 

To evaluate the effectiveness of AdaSpeech 2 for custom voice, we conduct experiments to train the source TTS model on LibriTTS dataset and adapt the model on VCTK and LJSpeech datasets with different adaptation settings. The average MOS of our results is $3.38$ on VCTK and $3.42$ on LJSpeech. The avarage SMOS of our results is $3.87$ while that of method using transcribed data is $3.96$, higher than other baseline. Besides, we also try to use our internal spontaneous ASR speech data to fine-tune the model. The experiments illustrate that AdaSpeech 2 achieves better adaptation quality than baseline methods and close to the transcribed upper bound (AdaSpeech).

\section{Method}
\label{sec:method}

In this section, we first introduce the overview of our method, and then describe the adaptation pipeline using untranscribed data in the following subsections.

\begin{figure}[h]
  \centering
  \includegraphics[totalheight=2.7in]{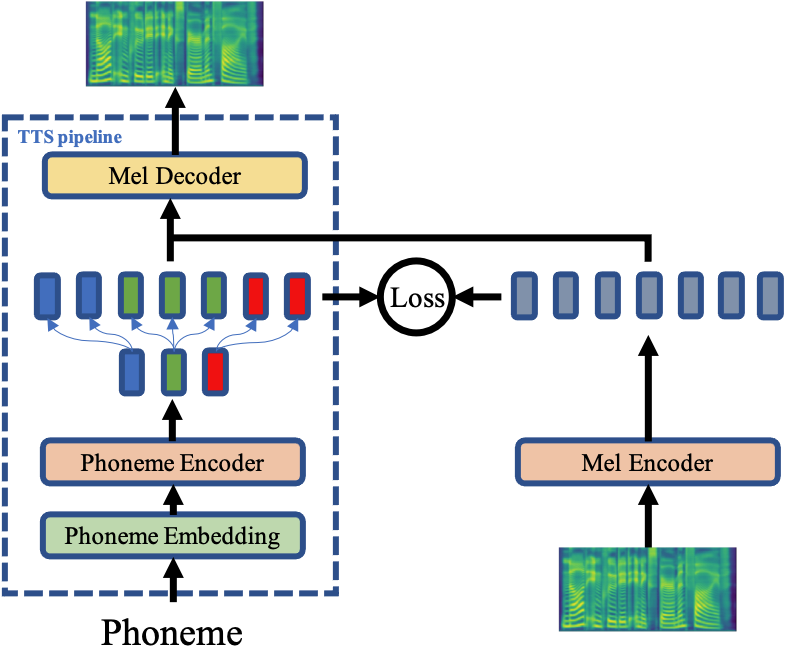}
  \caption{The model structure of AdaSpeech 2, where the TTS pipeline in the figure follows AdaSpeech~\cite{chen2021adaspeech}. Note that we also use the acoustic condition modeling and conditional layer normalization as in AdaSpeech, but do not show here mainly for simplicity.} \label{fig1}
\end{figure}

As shown in Figure~\ref{fig1}, our model has two main modules: 1) A TTS model pipeline with a phoneme encoder and a mel-spectrogram decoder, which is based on the backbone of AdaSpeech~\cite{chen2021adaspeech} considering it is one of the most popular non-autoregressive structure with fast and high-quality speech synthesis. 2) An additional mel-spectrogram encoder, which can leverage untranscribed speech data for adaptation.

\begin{figure}[tbh]
  \centering
  \includegraphics[totalheight=2in]{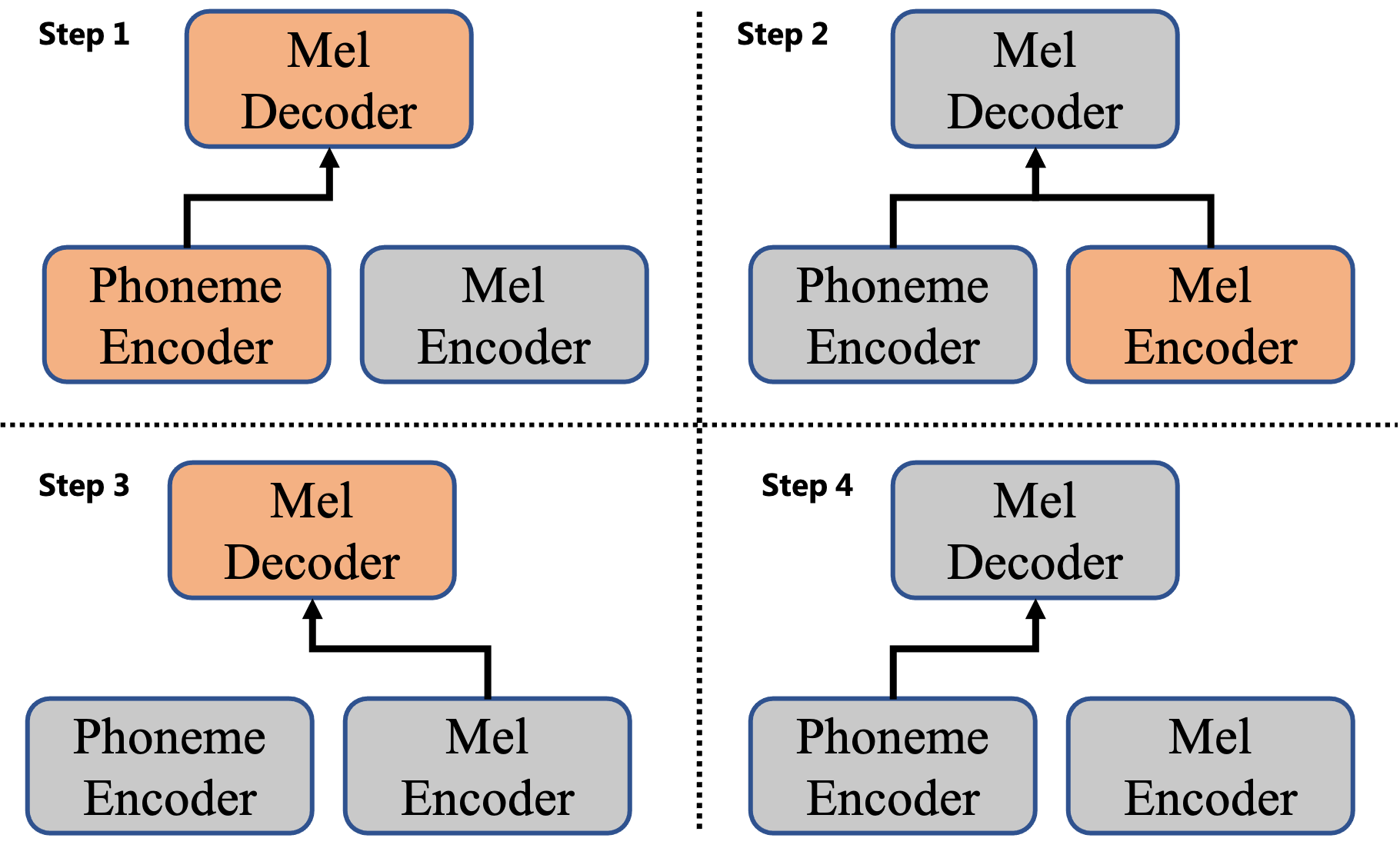}
  \caption{The four-step adaptation pipeline in AdaSpeech 2, where the first three steps represent the training process (source model training, mel encoder aligning and untranscribed speech adaptation) and the last step represents the inference process. Orange blocks represent the modules that are optimized while grey blocks represent the modules unchanged in the current step.} \label{fig2}
\end{figure}

As shown in Figure~\ref{fig2}, the adaptation pipeline consists of the following steps: 1) Source model training, where we need to train a multi-speaker TTS model and use it for adaptation. 2) Mel encoder aligning, where we need to train a mel encoder and use an alignment loss to make the output space of the mel encoder to be close to that of the phoneme encoder. 3) Untrancribed speech adaptation, where the mel decoder is adapted to the target speaker with the help of the mel encoder through auto-encoding on untranscribed data. 4) Inference, where we use the phoneme encoder and mel decoder to synthesize voice for the target speaker. In the following subsections, we introduce each step in details.

\subsection{Source Model Training}
\label{ssec:source}

Our TTS model pipeline follows the basic structure used in AdaSpeech~\cite{chen2021adaspeech}, with specifically designed acoustic condition modeling and conditional layer normalization for efficient TTS adaptation. The phoneme encoder consists of 4 feed-forward Transformer blocks~\cite{ren2020fastspeech,vaswani2017attention} and converts the phoneme sequence into the hidden sequence, which is further expanded according to the duration of each phoneme to match to the length of mel-spectrogram sequence.
The ground-truth duration is obtained from a forced-alignment tool~\cite{mcauliffe2017montreal}, and is used to train a duration predictor that built upon the phoneme encoder. The ground-truth duration is used for expansion in training while the predicted values are used in inference. The mel decoder consists of 4 feed-forward Transformer blocks and takes the expanded hidden sequence from the encoder as input. The decoder adds more variance information including pitch and more phoneme-level acoustic condition information as used in \cite{chen2021adaspeech}. The phoneme-level acoustic vectors are also expanded to the length of mel-spectrogram sequence and added into the decoder input with a dense layer.

\subsection{Mel-spectrogram Encoder Aligning}
\label{ssec:subhead}

We introduce an additional mel-spectrogram encoder that is designed to make use of untranscribed speech data through reconstruction (auto-encoding) for adaptation. From the perspective of the mel decoder, it expects the outputs of the phoneme and mel encoders are in the same space, so that the mel decoder adapted with speech reconstruction (taking the output of mel encoder as input) can be smoothly switched to the speech synthesis process (taking the output of the mel encoder) during inference.

After obtaining a well-trained TTS model as described in last subsection, we add this additional mel encoder and constrain the latent space obtained from the mel encoder to be close to that of phoneme encoder, by using the source transcribed speech data. The hidden sequence generated by the mel encoder is constrained with an L2 loss to match to the hidden sequence from the phoneme encoder, where two hidden sequences are of the same length since the phoneme hidden sequence has been expanded according to the phoneme duration. The mel encoder also consists of 4 feed-forward Transformer blocks, considering the symmetry of the system.

The alignment process above distinguishes our work from previous works~\cite{luong2020nautilus} that also adopt hidden sequence alignment to align the phoneme and mel encoder. We fix the parameters of the source TTS model (phoneme encoder and mel decoder) and only update the parameters of the mel encoder, which can form a plug-and-play way to leverage any well-trained source TTS model. On the contrary, previous works usually requires the training of source TTS model and adaptation process at the same time, which is not pluggable and may limit the broad usage in TTS adaptation. 

\vspace{-0.1cm}
\subsection{Untranscribed Speech Adaptation}
\label{sec:untranscribed}

The untranscribed data from the target speaker is used to fine-tune the model through the way of speech reconstruction with the mel encoder and decoder. To fine-tune as small amount of parameters as possible while ensure the adaptation quality, we adapt the parameters related to the conditional layer normalization following AdaSpeech~\cite{chen2021adaspeech}.

\vspace{-0.1cm}
\subsection{Inference}
\label{ssec:inference}

After the aligning and adaptation process mentioned above, the TTS pipeline can imitate the custom voice for speech synthesis. We use the original unadapted phoneme encoder and the partial adapted mel decoder to form the personalized TTS system to generate voice for each specific speaker.

\section{Experiments and Results}
\label{sec:typestyle}

\subsection{Datasets and Experimental Setup}
\label{ssec:Experimental Setup}

We train the source TTS model on LibriTTS~\cite{zen2019libritts} dataset and adapt on the VCTK~\cite{veaux2016superseded} and LJSpeech~\cite{ito2017lj} datasets. The LibriTTS dataset contains 586 hours speech data with 2456 speakers, while the VCTK dataset contains 44 hours speech data with 108 speakers. LJSpeech is a single-speaker high quality dataset with a total length of 24 hours.
Besides, we also use an internal speech dataset which is closer to daily conversation with spontaneous speech, to verify the performance of our method on challenging scenarios. 

We conduct the following preprocessings in the corpora: 1) Convert all the speech sampling rate to 16KHz; 2) Extract the mel-spectrogram with 12.5 ms hop size and 50ms window size following the common practice in~\cite{chen2021adaspeech}; 3) Convert text sequence into phoneme sequence with grapheme-to-phoneme conversion~\cite{sun2019token}.


The hidden dimension (including the embedding size, the hidden in self-attention, and the input and output of feed-forward network) is set to 256. The attention head, the feed-forward filter size and kernel size are set to 2, 1024 and 9 respectively. The output linear layer converts the 256-dimensional hidden into 80-dimensional mel-spectrogram. 

In source model training, first we take 100,000 steps to optimize the TTS pipeline. In mel-spectrogram encoder aligning and untranscribed speech adaptation, it takes much less steps (10,000) to train the additional mel encoder and 2,000 steps to adapt the mel decoder, which is much easier than re-training the entire model (that is why we call it pluggable). We train our model on 4 NVIDIA P40 GPUs and fine-tune it on 1 NVIDIA P40 GPU. 
Adam optimizer is used with $\beta_1=0.9$, $\beta_2=0.98$, $\epsilon=10^{-9}$.


\subsection{The Quality of Adaptation Voice}

We compare our proposed AdaSpeech 2 with other settings, which include: 1) GT, the ground-truth recordings; 2) GT mel + Vocoder, using ground-truth mel-spectrogram to synthesize waveform with MelGAN vocoder~\cite{kumar2019melgan}; 3) a joint-training method; 4) a PPG-based method; 5) AdaSpeech~\cite{chen2021adaspeech}. Synthesized speech samples can be found at \url{https://speechresearch.github.io/adaspeech2/}.

The joint-training method (Joint-training) trains the mel encoder and the phoneme encoder at the same time, which is similar to some previous adaptable TTS systems using untranscribed data~\cite{luong2020nautilus}. We take this method as a baseline to prove that the orderly training strategy in our method can effectively avoid the deviation of the output space of the encoders and improve the performance of adaptation.

The PPG-based method (PPG-based) uses PPGs (phonetic posteriorgrams)~\cite{sun2016personalized, saito2018non} extracted from the untranscribed speech to fine-tune the TTS model. In this setting, we replace the mel-spectrogram encoder in our proposed pipeline with an additional PPG encoder, which has the same structure but takes PPG sequence as input. The PPG sequence is extracted by an internal production ASR model, with a hidden dimension of 512 and the same length of the mel-spectrogram sequence. It is transformed into 256 dimension with a dense layer and fed into the PPG encoder. We take the performance of the PPG-based method as an upper bound since it introduces similar text/phoneme information from an additional ASR system.

AdaSpeech~\cite{chen2021adaspeech} is a previous TTS adaptation system using paired text and speech (transcribed speech) data. We take its performance as another upper bound.

To make them comparable, we use the same encoder and decoder structure and fine-tune the same parameters in these three baseline methods. We randomly choose 6 speakers (including 3 men and 3 women) from VCTK and the only single speaker from LJSpeech for adaptation.
We take 50 sentences for adaptation and synthesize 15 sentences for each target custom voice.
Twenty native English speakers are asked to make quality judgments in terms of naturalness and similarity.

\begin{table}[tbh]
  \centering
  \caption{The MOS and SMOS with $95\%$ confidence intervals when adapting the source TTS model to VCTK and LJSpeech. MelGAN is selected as the vocoder in each case.} \label{tab1}
  \vspace{-0.3cm}
  \begin{tabular}{l | l | c | c}
    \toprule
    \textbf{Metric} & \textbf{Setting} & \textbf{VCTK} & \textbf{LJSpeech}\\
    \midrule
    \multirow{5}{*}{\small MOS} 
    & \textit{GT} & $3.58\pm 0.12$ & $3.63\pm 0.11$ \\
    & \textit{GT mel+Vocoder} & $3.42\pm 0.12$ & $3.49\pm0.11$\\
    & \textit{Joint-training} &  $2.91\pm 0.09$ & $2.89\pm
    0.12$\\
    & \textit{PPG-based} &  $3.39\pm 0.11$ & $3.44\pm 0.12$\\
    & \textit{AdaSpeech} &  $3.39\pm 0.10$ & $3.45\pm 0.11$\\
    \cmidrule{2-4}
    & \textit{AdaSpeech 2} & $3.38\pm 0.12$ & $3.42\pm 0.12$ \\
    \midrule
    \midrule
    \multirow{5}{*}{\small SMOS}
    & \textit{GT} & $4.20\pm 0.12$ & $4.24 \pm 0.09$ \\
    & \textit{GT mel+Vocoder} & $4.06\pm 0.08$ &$4.02\pm 0.11$\\
    & \textit{Joint-training} & $3.71\pm 0.13$ &$3.19\pm 0.16$\\
    & \textit{PPG-based} & $3.82\pm 0.11$ &$3.51\pm 0.15$\\
    & \textit{AdaSpeech} & $3.94\pm 0.12$ &$3.59\pm 0.12$\\
    \cmidrule{2-4}
    & \textit{AdaSpeech 2} & $3.84\pm 0.08$& $3.51\pm 0.12 $\\
    \bottomrule
  \end{tabular}
\end{table}

\vspace{-0.3cm}

We evaluate the MOS (Mean Opinion Score) and SMOS (Similarity MOS) of the generated audio samples. The results are shown in Table~\ref{tab1}. We have several observations: 1) Our model achieves on par voice quality with ground-truth recording and the two upper bounds (AdaSpeech \cite{chen2021adaspeech} and the PPG-based method), and is superior to the joint-training method. 2) The performance of our model on SMOS is slightly worse than the upper bound method (about $0.1$ lower than AdaSpeech), but still better than the joint-training method. 
We also test CMOS (comparison MOS) between our model and the transcribed speech adaptation (AdaSpeech) on our internal spontaneous speech data and achieve on-par performance (AdaSpeech achieves only $0.012$ CMOS score higher than our model). These results demonstrate the effectiveness of our proposed adaptation strategy.

\subsection{Analyses on Adaptation Strategy}

In this section, we design two experiments to demonstrate the effectiveness of the adaptation strategy. 

First, we conduct an ablation study to verify the effectiveness of aligning the hidden output space of the mel encoder to that of the phoneme encoder by an L2 loss. In this setting, the mel encoder is trained without the restrict of the L2 loss (other losses remain) in the step of mel encoder aligning. Then we evaluate the CMOS between this  setting and our proposed method. The results shown in Table~\ref{tab2} demonstrate the necessity of using this loss constraint. 

To justify the method of only adapting the mel decoder to ensure the consistency of the output space of the phoneme encoder and the mel encoder, we conduct another experiment to fine-tune the whole mel encoder together with the mel decoder in adaptation. As shown in Table~\ref{tab2}, adapting the mel encoder leads to the decrease in voice quality. It can be seen that keeping the encoders unchanged indeed have a positive effect on the adaptation performance.

\begin{table}[tbh]
  \centering
  \caption{Analyses on the adaptation strategy. \textit{Origin} represents the original pipeline that constrains the mel encoder and the phoneme encoder with an L2 loss, and only fine-tune a part of parameters in the mel decoder.}
  \vspace{-0.3cm}
  \label{tab2}
  \begin{tabular}{ l | c | c}
    \toprule
    \textbf{Setting} & \textbf{CMOS} & \textbf{SMOS}\\
    \midrule
    \textit{Origin} & $0$ & $3.89\pm 0.12$\\
    \midrule
    \textit{Without L2 loss constraint} & $-0.112$ & $3.82\pm 0.12$\\
    \textit{Fine-tune mel encoder \& decoder} & $-0.132$ & $3.79 \pm 0.12$\\
    \bottomrule
  \end{tabular}
\end{table}


\vspace{-0.3cm}
\subsection{Varying Adaptation Data}


\begin{wrapfigure}{r}{0pt}
    \centering
    \includegraphics[width=0.26\textwidth]{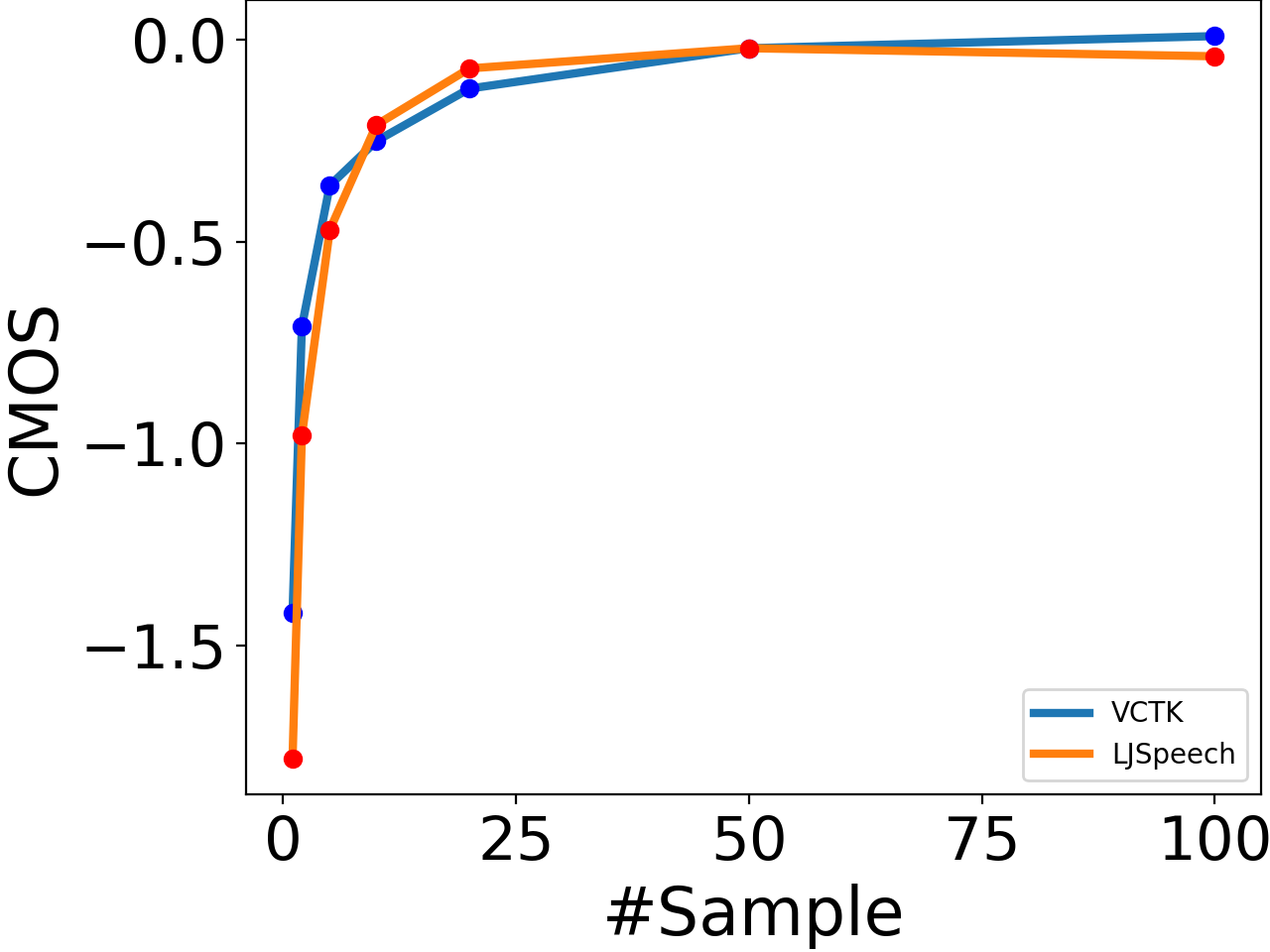}
    \caption{The CMOS of different adaptation data on LJSpeech and VCTK. Ground-truth is taken as the comparison system in CMOS test and the CMOS is set to be $0$.} \label{fig3}
\end{wrapfigure}

We conduct CMOS evaluation when using different amount of adaptation data (1, 2, 5, 10, 20, 50, 100 samples respectively) on VCTK and LJSpeech. As shown in Figure~\ref{fig3}, when the adaptation data is less than 20, the adaptation quality is significantly improved with the increase of data. If we continue to increase the amount of data from that, the improvement is not obvious.

\section{Conclusion}
\label{sec:Conclusion}

In this paper, we develop AdaSpeech 2, a pluggable and effective adaptive TTS system using untranscribed speech data. We introduce an additional mel-spectrogram encoder to a source TTS pipeline and constrain the output sequence of the speech encoder to be close to that of the TTS phoneme encoder, without adjusting the source TTS pipeline. We present the pipeline of source model training, mel-spectrogram encoder aligning, untranscribed speech adaptation and inference for custom voice. We achieve close or on par voice quality and similarity with the upper bound systems (including AdaSpeech~\cite{chen2021adaspeech}), in terms of MOS and SMOS. For future work, we will explore different adaptation methods to improve voice quality and similarity and further extend our method to more challenging scenarios such as spontaneous speech.

\bibliographystyle{IEEEbib}
\bibliography{main}

\end{document}